\documentclass[12pt]{article}

\usepackage{epsfig}

\def\beq{\begin{equation}}
\def\eeq{\end{equation}}
\def\bea{\begin{eqnarray}}
\def\beaa{\begin{eqnarray*}}
\def\eea{\end{eqnarray}}
\def\eeaa{\end{eqnarray*}}
\def\bq{\begin{quote}}
\def\eq{\end{quote}}
\def\gappeq{\mathrel{\rlap {\raise.5ex\hbox{$>$}}
{\lower.5ex\hbox{$\sim$}}}}
\def\lappeq{\mathrel{\rlap{\raise.5ex\hbox{$<$}}
{\lower.5ex\hbox{$\sim$}}}}

{\end{list}}
\newcounter{enumct}

\newcommand{\captive}[1]{\rule{5mm}{0mm}%
\begin{minipage}{150mm}\caption[small]{#1}\end{minipage}}
 
\begin{document}
 
%set sloppy attitude to line breaks
\sloppy

%do not number pages
\pagestyle{empty}

\rightline{CERN-TH/2000-050}
\rightline{hep-ph/0002117}
\begin{center}
{\LARGE\bf Physics Goals of the Next Century $@$ CERN }
\\[10mm]
{\Large John Ellis} \\[3mm]
{\it Theoretical Physics Division, CERN}\\[1mm]
{\it CH - 1211 Geneva 23}\\[1mm]
{\it E-mail: john.ellis@cern.ch}\\[20mm]

{\bf Abstract}\\[1mm]
\begin{minipage}[t]{140mm}
After recalling briefly the main physics issues beyond the Standard Model,
the main physics objectives of experiments at CERN in the coming
decade(s) are reviewed. These include the conclusion of the LEP
programme during the year 2000, a limited number of fixed-target
experiments during the following years, the CNGS long-baseline
neutrino programme and the LHC, both scheduled to start in 2005.
Then possible accelerator
projects at CERN after the LHC are reviewed, in the expectation
that an $e^+ e^-$ linear collider in the TeV energy range will be built
elsewhere. The default
option for CERN's next major project may be the CLIC multi-TeV $e^+ e^-$
collider project. Also interesting is the option of a three-step scenario
for muon storage rings, starting with a neutrino factory, continuing with
one or more Higgs factories, and culminating in a $\mu^+ \mu^-$ collider
at the high-energy frontier. \end{minipage}\\[5mm]

\rule{160mm}{0.4mm}
{\it To appear in the \\
Proceedings of the
Fifth International Conference on the \\
Physics Potential And Development of Muon Colliders, \\
15 - 17 Dec. 1999, San Francisco, California}

\end{center}
\vfill\eject
\pagestyle{plain}
\setcounter{page}{1}
\section{Beyond the Standard Model}

As at other accelerator laboratories, the top priorities at CERN in the 21st
century will be experiments probing beyond the Standard Model. Indeed, this is
surely the only responsible motivation for major new accelerators.

Problems beyond the Standard Model may conveniently be gathered into three major
classes: those of {\bf Mass}, {\bf Unification} and {\bf Flavour}. What is the
origin of the particle masses, are they due to a Higgs boson, and if so why are
they so small, perhaps because of supersymmetry? Is there a simple group framework
containing the strong, weak and electromagnetic gauge interactions, and does it
predict new observable phenomena such as proton decay and neutrino masses? Why are
there so many types of quarks and leptons, and how can one understand their weak
mixing and CP violation, perhaps because they are composite or have extra
symmetries?

Beyond these beyonds lurks the {\bf Theory of Everything} that is supposed also to
include gravity, reconcile it with quantum mechanics, explain the origin of
four-dimensional space-time, etc. The only plausible candidate is what used to be
called superstring theory, now M theory. There are ideas how to probe this at
accelerators, particularly if there are large extra dimensions, but these will not
be featured in the rest of this talk.

\section{Setting The Stage}

By comparison with other high-energy physics laboratories, CERN is fortunate to
have an exciting physics programme beyond the year 2005 already approved
and under
construction, centred on the LHC. However, the time scales for the R\&D,
approval
and construction of major new accelerators are very long: the first LEP physics
study started in 1975~\cite{LEPYB}, 14 years before the first data, and
the first LHC
physics
study was in 1984~\cite{Lausanne}. Therefore, it is already time to be
thinking what CERN might
do for an encore after (say) ten years of physics with the LHC. Although
necessary, extrapolation to the likely physics agenda beyond 2015 is foolhardy,
since several major accelerators will be providing cutting-edge data during the
intervening period, and we do not know what they will find. (Otherwise, it
would
not be research, would it?) Nevertheless, we should try to set the apr\`es-LHC
era in context by surveying the ground that these intervening accelerators will
cover~\cite{EKR}, even if our crystal ball does not reveal what they will
find there.

LEP operation will terminate in 2000, after providing sensitivity to Higgs
masses
below about 110 GeV. The current lower limit from the  data of an individual
LEP experiment reaches about 106~GeV, as seen in Fig. 1~\cite{LEPC},
and a combined analysis of the full 1999 data might
increase the sensitivity to about 109 GeV. The most optimistic projection for
2000 that I have seen would extend this to about 113 GeV. Clearly, the
overall
picture changes if LEP discovers the Higgs boson. However, the precision
electroweak data and supersymmetric models independently suggest that $m_H
\lappeq$ 200 GeV, as seen in Fig. 2~\cite{LEPEWWG}, in which case the
programme of
exploring in
detail the properties of the Higgs boson is already well posed, just as the LEP
programme was outlined before the discovery of the $W^\pm$ and $Z^0$.

\begin{figure}%1
\begin{center}
\mbox{\epsfig{file=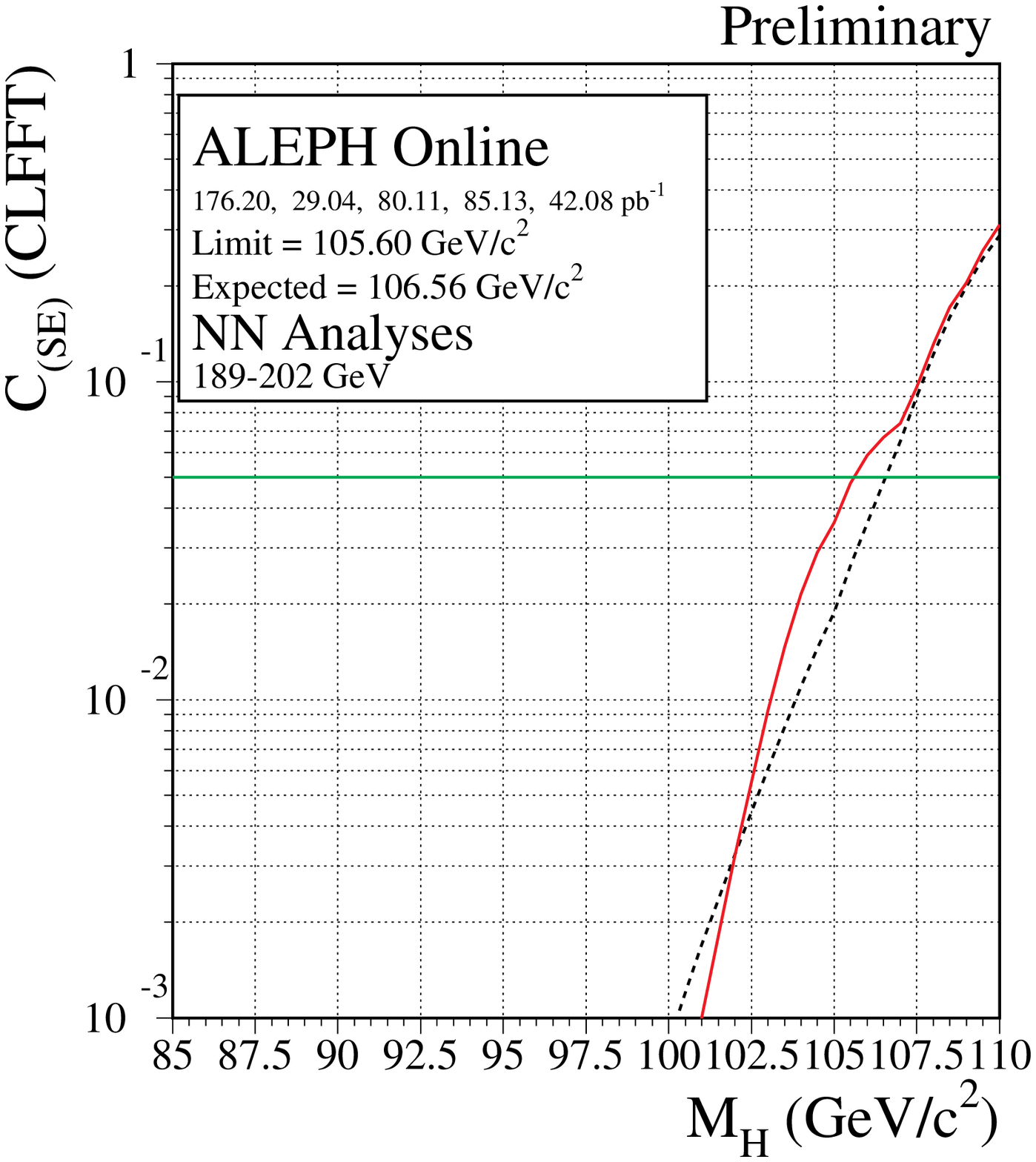,width=79mm}}
%\mbox{\epsfig{file=fig1b.eps,width=79mm}}
\end{center}
\captive{\it Preliminary lower limit on the Standard Model Higgs mass
obtained by the ALEPH collaboration~\cite{LEPC,ALEPH}.}
\label{figure}
\end{figure}

\begin{figure}%2
\begin{center}
\mbox{\epsfig{file=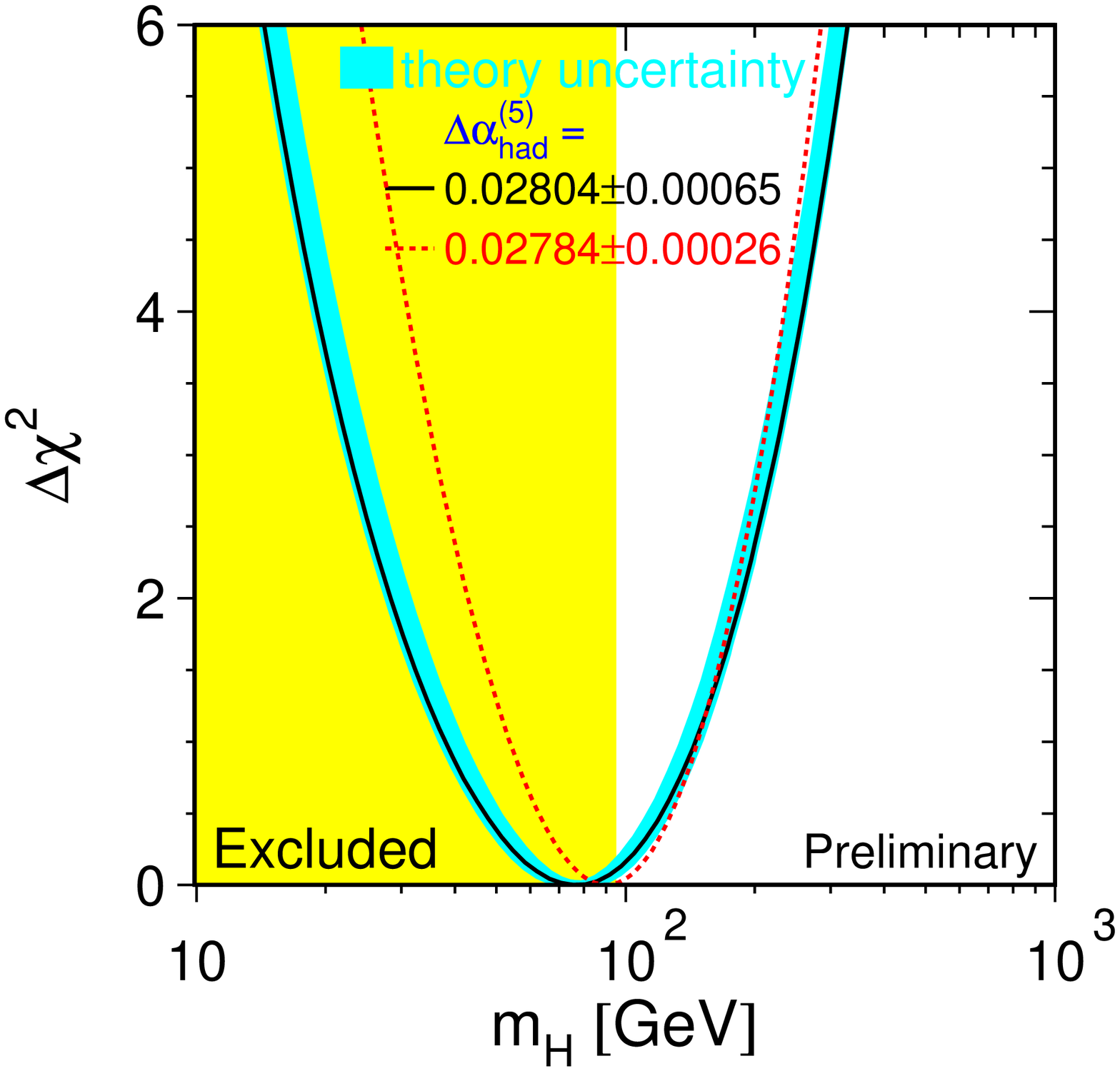,width=8cm}}
\end{center}
\captive{\it Estimate of the Higgs boson mass obtained from precision
electroweak data~\cite{LEPEWWG}.}
\end{figure}

CDF and
D$\phi$  have a
chance to find the Higgs
boson before the LHC in the next run of the FNAL Tevatron collider 
starting in 2001, as seen in Fig.~3~\cite{Carena}. This
figure is based on theoretical assessments of the capabilities of the Tevatron
detectors, and the experiments may fare better or worse. However, taken at face
value, it seems that the Tevatron detectors would need more than 5 or even 10
pb$^{-1}$ to explore masses beyond LEP's reach. Will these be available for the
LHC's scheduled start in 2005? FNAL's window of opportunity will extend
somewhat
beyond LHC start-up, since ATLAS and CMS  will take some time  to accumulate
the luminosity needed to explore the difficult region $M_H \lappeq$ 130
GeV~\cite{TDR}.

\begin{figure}%3
\begin{center}
\mbox{\epsfig{file=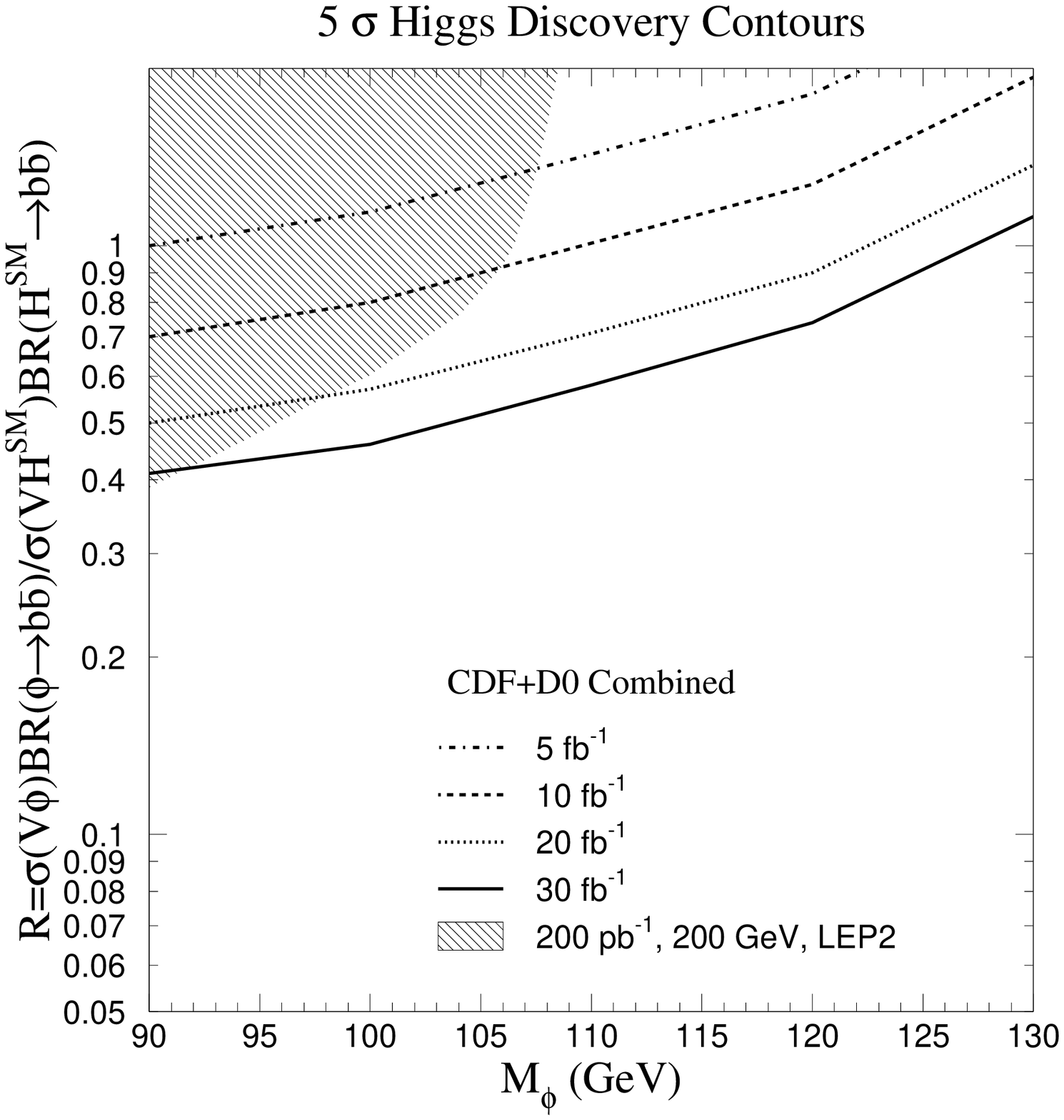,width=8cm}}
\end{center}
 \captive{\it Comparison of the estimated physics reaches for
Higgs searches at LEP~2 and the FNAL Tevatron collider~\cite{Carena},
as a function of Higgs mass and collider luminosity.}
\end{figure}

The CERN experimental programme in the years leading up to LHC operation has been
pared to the bone, because of resource restrictions. The NA48 apparatus will
continue to be used to measure $\epsilon^\prime/\epsilon$ and rare K decays,
probably through the year 2003. The COMPASS experiment is starting its 
programme of
gluon polarization 
and other measurements. The DIRAC experiment has started its 
programme to
measure $\pi - \pi$ scattering lengths via the lifetime of the pionium atom. The
antiproton-decelerator (AD) facility is being commissioned for its programme of
anti-Hydrogen spectroscopy and CPT tests. The neutron time-of-flight facility and
ISOLDE will be in operation for nuclear physics. The SPS heavy-ion programme, which
has found evidence for collective effects indicative of a new state of 
matter that may be associated with the
quark-gluon plasma, may acquire a new lease of life from a charm production
experiment~\cite{ELFE}. The HARP experiment~\cite{HARP} is being proposed
to study
particle production for
$\nu$ factory designs and to help reduce uncertainties in the atmospheric $\nu$
flux. Finally, a proposal is being prepared to study the seeding of clouds
by ionizing particles~\cite{CLOUD}, with the aim of seeing whether
fluctuations in
the flux of cosmic rays could influence the amount of cloud cover, and
hence the climate~\cite{Nature}.

\section{The CERN-Gran Sasso Long-Baseline $\nu$ Project}

A promising new area of exploration has been opened by the strong
indications for
neutrino oscillations found by Super-Kamiokande~\cite{SK} et
al.~\cite{al}, and several
long-baseline
neutrino projects are underway. K2K has started taking data, and will be
able to
measure $\nu_\mu$ disappearance in much of the region of atmospheric-neutrino
parameter space favoured by Super-Kamiokande~\cite{K2K}. The data
taken in 1999 already find intriguingly few events, compared with
no-oscillation expectations. Starting
in 2001, KamLAND~\cite{KamLAND} will
explore the large-mixing-angle (LMA) MSW solution of the solar-neutrino
problem.
In 2003/2004, MINOS will start exploring $\nu_\mu$ disappearance, the NC/CC
ratio and other oscillation signatures in the FNAL NuMI beam~\cite{MINOS}.

We heard at this meeting that CERN-Gran Sasso neutrino beam project
(CNGS)~\cite{CNGS} has been
approved by the CERN Council. The favoured interpretation of the Super-Kamiokande
and other data is $\nu_\mu\rightarrow\nu_\tau$ oscillations, but the K2K and
NuMI/MINOS projects are not designed to look directly for $\tau$ appearance: this
is the primary objective of the CNGS project. I believe that direct observation of
the $\tau$ oscillation product is an important scientific objective: ``If you have
not seen the body, you have not proven the crime", cf. Jimmy Hoffa and the
discovery of the gluon in $e^+e^-\rightarrow$ three-jet events~\cite{EGR}.

The CNGS project makes maximal use of the CERN and Gran Sasso  infrastructure,
including the SPS-LHC beam transfer line and the preplanned orientation of the Gran
Sasso experimental halls. The CNGS beam energy has been optimized for $\tau$
production, and there are two major experiments proposed
for the CGNS beam: OPERA~\cite{OPERA} and ICANOE~\cite{ICANOE}. These
expect to detect the following numbers of $\tau$
(background) events if $\Delta m^2 = 3.5 \times 10^{-3}$ eV$^2$: 18 ($<$1) for OPERA,
44(6.4) for ICANOE. OPERA will be able to see $\tau$ production
 comfortably over the parameter region
favoured by Super-Kamiokande, as seen in Fig.~4, and ICANOE may
additionally be
able to probe the LMA MSW solution of the solar-neutrino problem, via low-threshold
measurements of atmospheric neutrinos, as seen in Fig.~5.

\begin{figure}%4
\begin{center}
\mbox{\epsfig{file=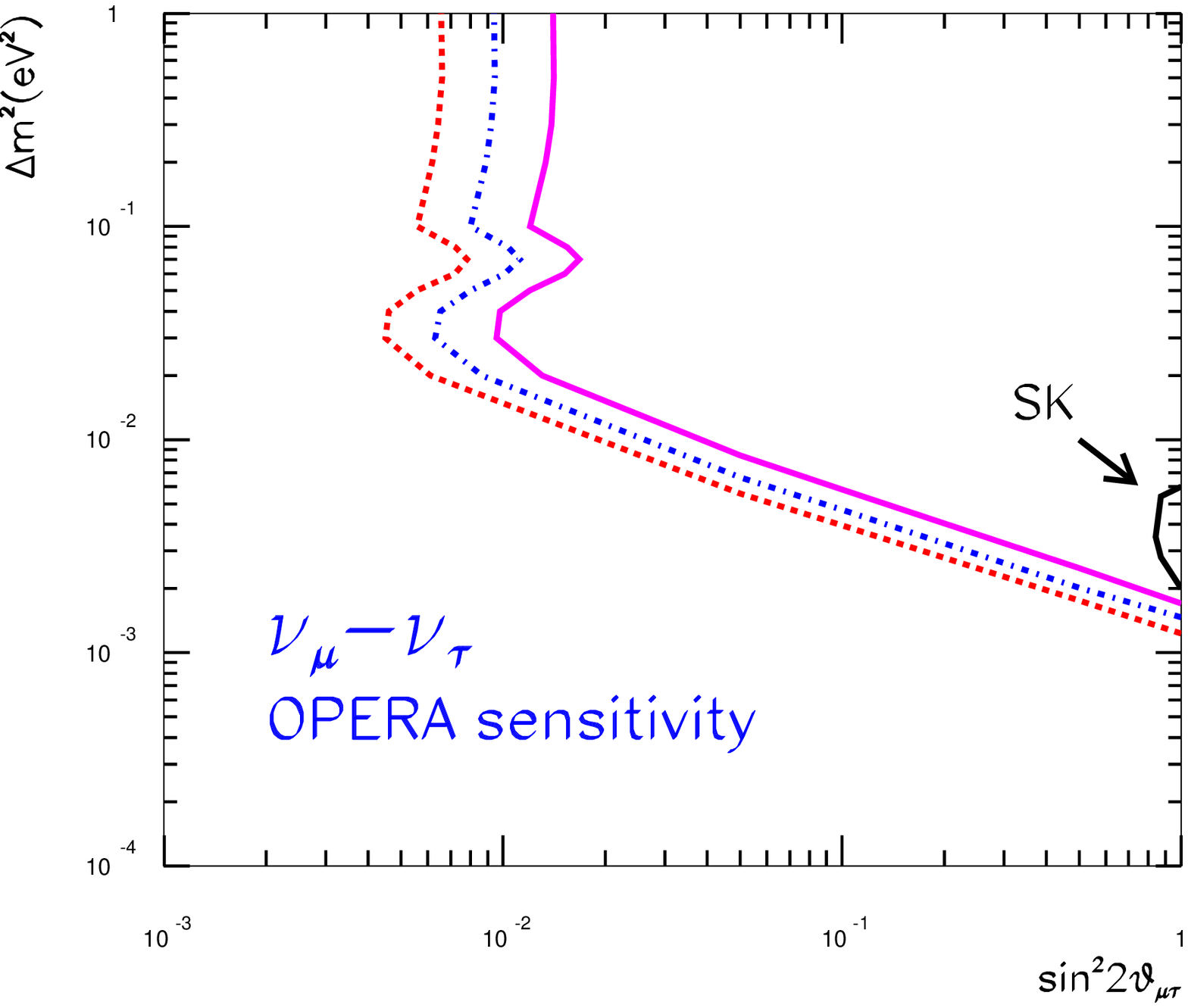,width=8cm}}
\end{center}
 \captive{\it Estimated sensitivity of the OPERA experiment
to $\tau$ production by $\nu_\mu \rightarrow \nu_\tau$
oscillations~\cite{OPERA}, after 2,3 and 4 years of operation,
compared with the region preferred by Super-Kamiokande~\cite{SK}.}
\end{figure}

\begin{figure}%5
\begin{center}
\mbox{\epsfig{file=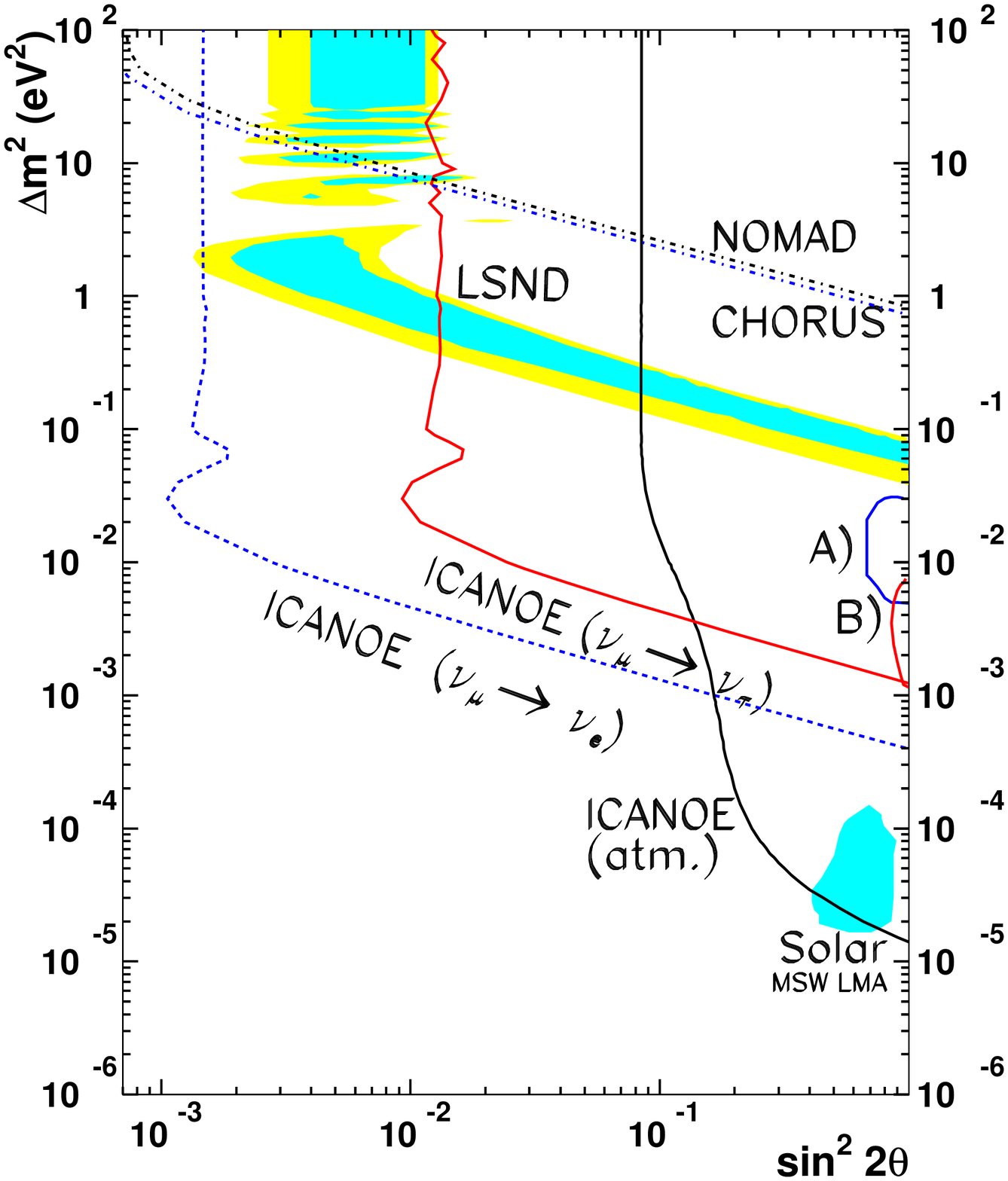,width=8cm}}
\end{center}
 \captive{\it Estimated sensitivity of the ICANOE experiment
to $\tau$ production by $\nu_\mu \rightarrow \nu_\tau$
oscillations, and to the LMA MSW solution to the solar
neutrino problem~\cite{ICANOE}.}
\end{figure}

\section{Status of LHC Construction}

As far as the LHC machine is concerned, the most critical items are the civil
engineering and the overall cost. The surface construction work is currently on
schedule. The ATLAS pit has been dug as far as possible down towards the LEP/LHC
tunnel. There have been some delays here and for the CMS pit, where there was more
difficulty than expected in stopping the underground water flow. There has also
been some delay in digging the TI8 (Eastern) SPS-LHC beam-transfer tunnel, but this
has no consequence for the overall schedule.  The TI2 (Western) SPS-LHC tunnel is
on the critical path, and the prospect of a delay there has led to a fall-back plan
to install LHC magnets through the L3/ALICE pit.

The second pre-production main-dipole magnet performs nominally, and there is hope
of operating the ring at 9 Tesla, which would correspond to E$_{\rm cm}$ = 15 TeV.
The contracts for the magnet cold masses and other components have been placed,
within budget, and the final assembly contracts will be placed in 2001. The various
international contributions to the machine, from the U.S., Japan, Russia, Canada
and India are proceeding well. For example, several trucks arrive at  CERN each
month delivering beam-transfer magnets from Novosibirsk. 

\section{Selected LHC Physics Topics}

Although the main lines of the LHC physics programme are well known, there are
continual advances. Here I highlight new aspects from the (published) ATLAS physics
TDR and the (forthcoming) CMS physics TDR.

As seen in Fig.~6, the LHC will discover the Standard-Model Higgs
boson (if this has not been done already), but this may take some
time~\cite{TDR}. 
However, for any given value of the Higgs mass, 
the LHC will probably measure only
one or two of its decay modes. 
It will be able to measure their branching ratios
with an accuracy of 10 to 20 \%, 
the total width with an error $\lappeq$ 10 \% for
$M_H \gappeq$ 300 GeV, and the mass with an error between $10^{-3}$ and $10^{-2}$
for $m_H \lappeq$ 800 GeV~\cite{ZKNR}.
The LHC will
also be able to discover Higgs bosons in the minimal supersymmetric
extension of
the Standard Model (MSSM), as seen in Fig.~7, though perhaps not all of
them. 

Over most of the MSSM Higgs parameter space, it will discover the Higgs in two or
more ways.
It will also find
supersymmetry (if this has not been done already), establish much of the
sparticle
spectrum, as displayed in Table 1~\cite{LHCsusy}, and measure some
distinctive spectral
features, as seen in Fig.~8. 
These will enable the MSSM parameters to be measured with good precision, at least
if universal input supergravity parameters are assumed. The cascade decays may also
be a rich source of MSSM Higgs bosons, as seen in Fig.~9~\cite{TDR}. It is
also worth
noting that the LHC can cover comfortably all the region of MSSM parameter space
where the lightest supersymmetric particle could constitute the cold dark matter in
the Universe. The LHC can also find supersymmetry in scenarios with R
violation~\cite{Baer} or gauge mediation~\cite{gaugemed}.

To baseline the subsequent discussion, we surmise
that the LHC will not only discover the Higgs boson, but also measure its mass
with a precision between 0.1 \% and 1 \%~\cite{TDR}. However, it will only
be able to
observe a couple of Higgs decay modes. Within the context of the MSSM, the LHC
will have found many sparticles, but perhaps not the heavier Higgs bosons and
weakly-interacting sparticles such as sleptons and
charginos~\cite{LHCsusy}. The spectroscopic
measurements will not enable the underlying MSSM parameters to be strongly
over-constrained.

\begin{table}[h]
\caption{\it The LHC as `Bevatrino': Sparticles detectable~\cite{LHCsusy}
at
five selected points in supersymmetric parameter space are denoted by +}
\begin{center}
\begin{tabular}{|c|c|c|c|c|c|c|c|c|c|c|c|c|}   \hline
&&&&&&&&&&&& \\
 & $h$ & $H/A$ & $\chi^0_2$ & $\chi^0_3$ & $\chi^-_1$  &$\chi^\pm_1$ & $\chi^\pm_2$ &
$\tilde q$ & $\tilde b$ & $\tilde t$ & $\tilde g$ & $\tilde\ell$\\ \hline 
%&&&&&&&&&&&& \\ 
1 & + && + &&&&& + & + & + & + &\\  \hline
2 & + &&+ &&&&& + & + & + & + & \\  \hline
3 & + & + & + &&&+ && + & + && + &\\  \hline
4 & + && + & + & + & + & + & + &&& + & \\ \hline
5 & + && + &&&&& + & + & + & + & +  \\ \hline
\end{tabular}
\end{center}
\end{table}

\begin{figure}%6
\begin{center}
\mbox{\epsfig{file=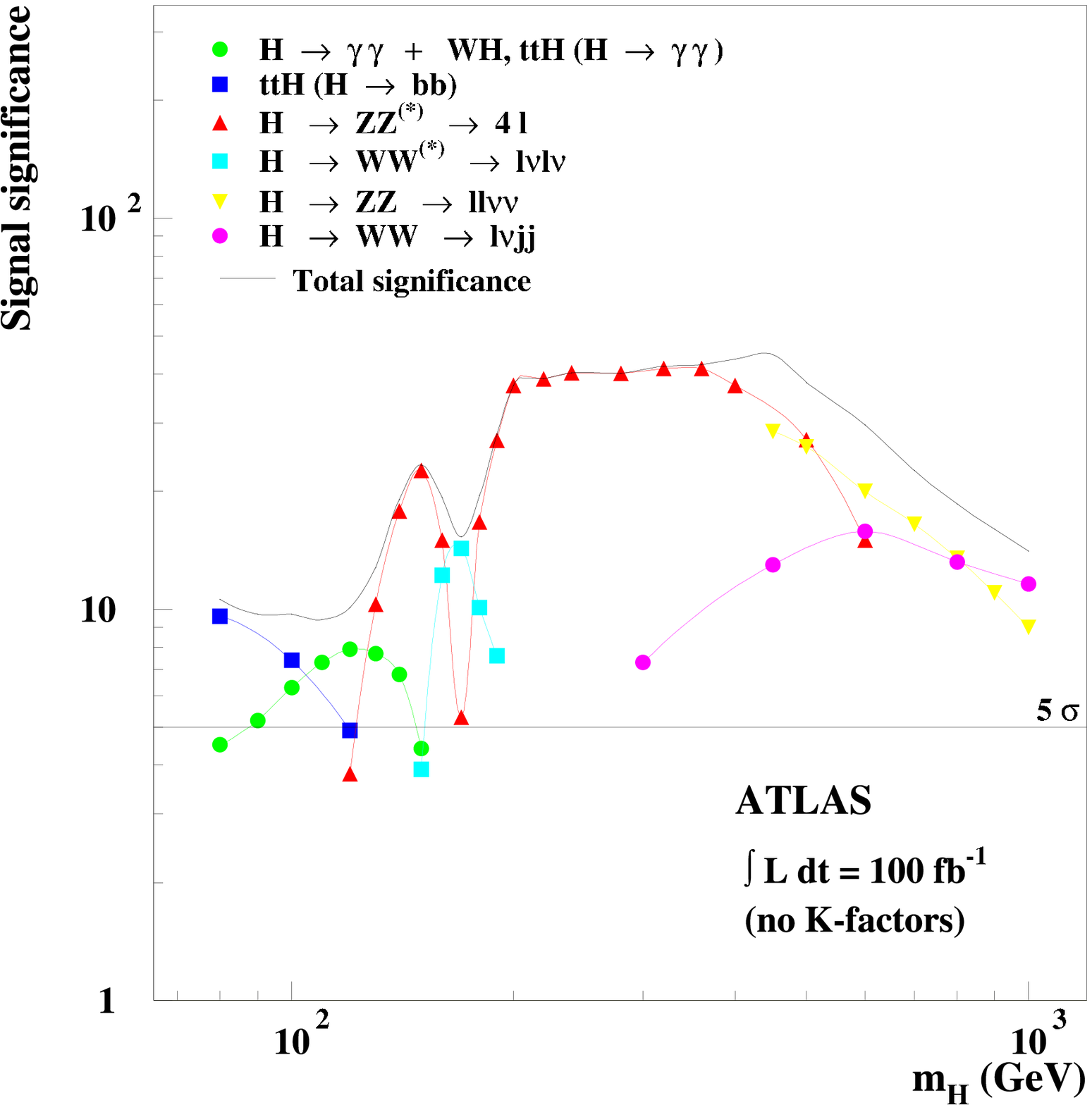,width=8cm}}
\end{center}
 \captive{\it Estimated significance of the possible
Higgs detection at the LHC in various channels, as
a function of the assumed value of the
Higgs mass~\cite{TDR}.}
\end{figure}

\begin{figure}%7
\begin{center}
\mbox{\epsfig{file=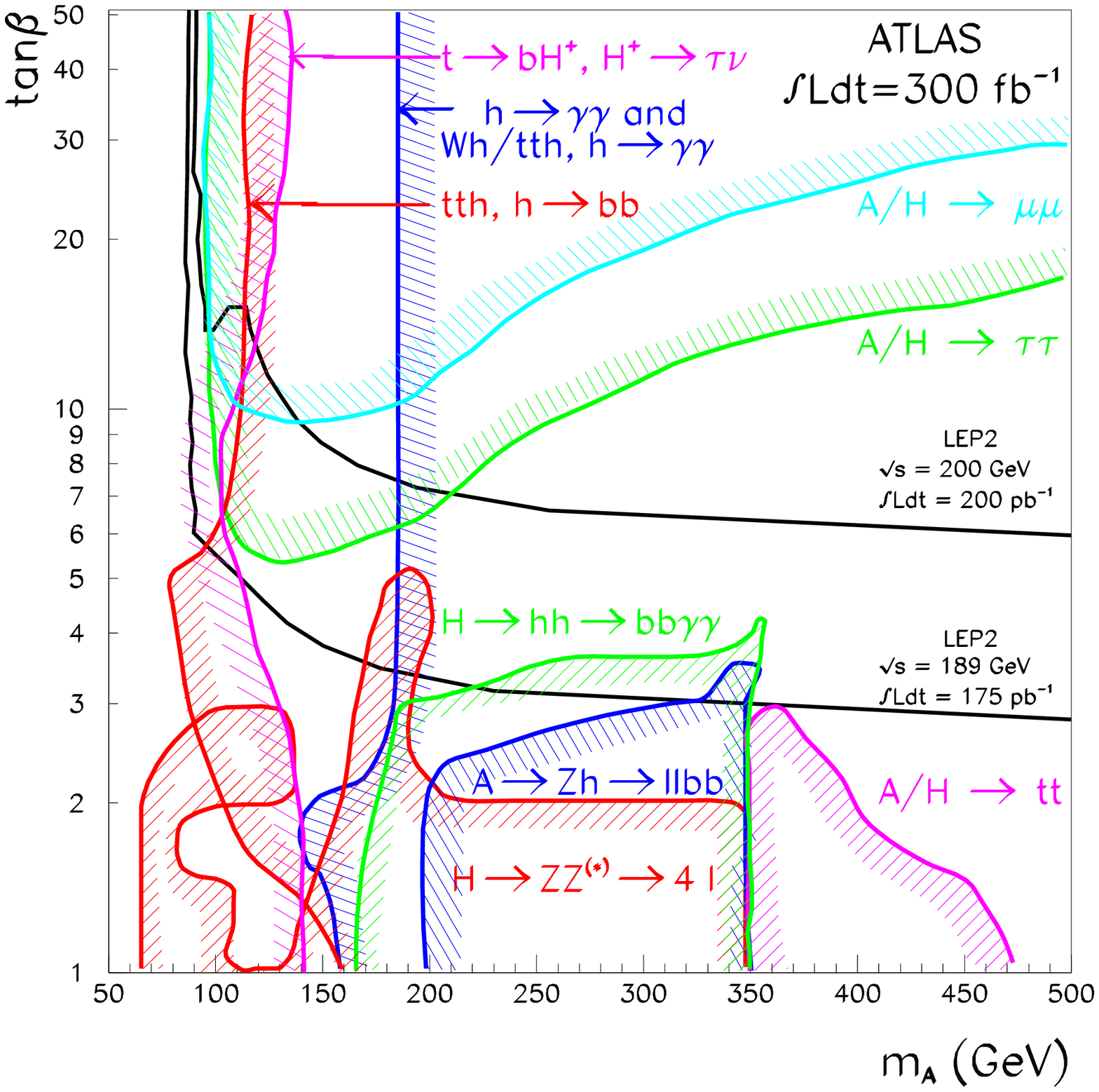,width=8cm}}
\end{center}   
 \captive{\it Detectability of MSSM Higgs bosons at the LHC~\cite{TDR}.}
\end{figure}

\begin{figure}%8
\begin{center}
\mbox{\epsfig{file=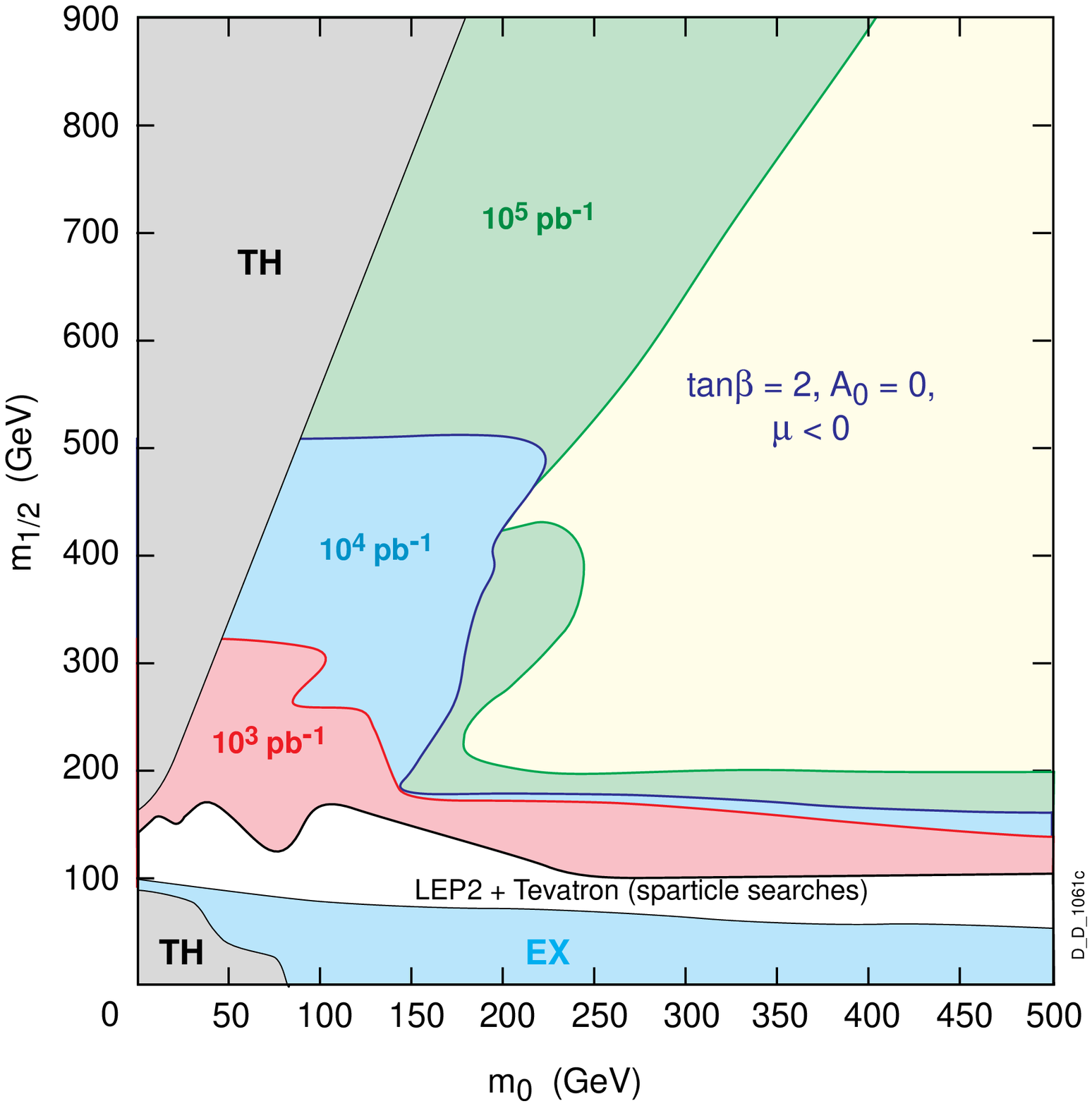,width=8cm}}
\end{center}
 \captive{\it Region of the MSSM parameter space in which
one can detect distinctive `edge' features in the dilepton spectra
due to cascade decays of sparticles, such as $\tilde q \rightarrow q
+ \chi_2, \chi_2 \rightarrow \chi_1 + \ell^+ \ell^-$~\cite{TDR},
for the indicated integrated LHC luminosities.}
\end{figure}

\begin{figure}%9
\begin{center}
\mbox{\epsfig{file=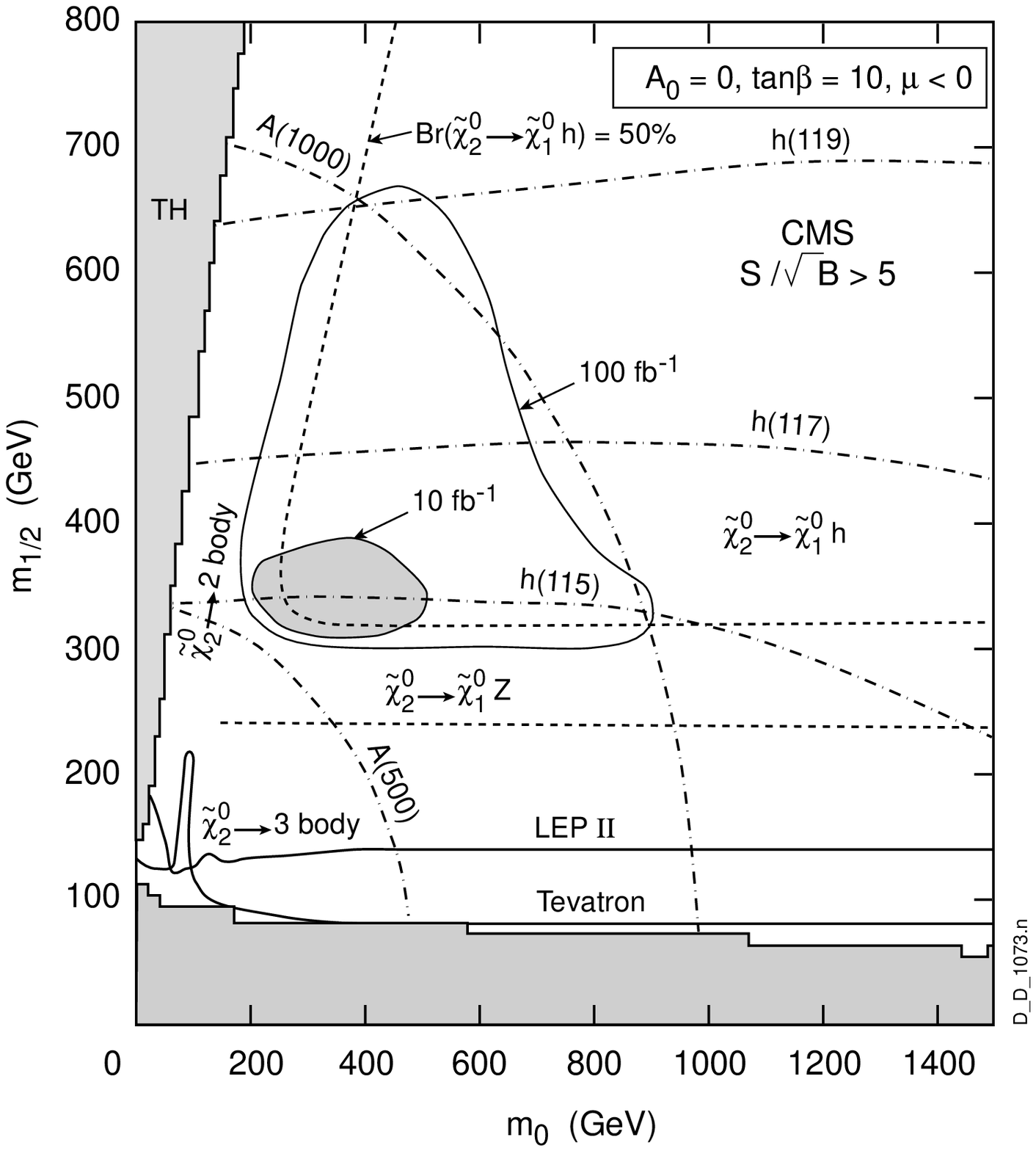,width=8cm}}
\end{center}
 \captive{\it Region of the MSSM parameter space in which
one can detect a Higgs boson in the
cascade decays of sparticles~\cite{TDR}, assuming universal input
supergravity parameters, for the indicated integrated LHC luminosities.}
\end{figure}

\section{$e^+e^-$ Linear-Collider Physics}

The stage is now set for the
entry of
the next major actor,
the
first-generation
$e^+e^-$ linear collider which is being proposed by other accelerator laboratories.
It will boast a very clean experimental
environment and
egalitarian production of new weakly-interacting
sparticles~\cite{DESY-ECFA}. Polarization
will be
a useful analysis tool, and $e\gamma, \gamma\gamma$ and $e^-e^-$ colliders will
come `for free'. In many ways, it will be complementary to the LHC. The
trickiest
issue may be how to fix its maximum energy scale. The location of the $\bar tt$
threshold is known, and the precision electroweak data~\cite{LEPEWWG} 
indicate that the ZH
threshold is probably below 300 GeV. This is also expected on the basis
of calculations of the lightest Higgs mass in the MSSM~\cite{HMSSM}.
However, what is the sparticle threshold
(assuming there is one), and how/when will be able to fix it? Flexibility
in the
linear-collider centre-of-mass energy is surely essential. In addition to the
$\bar tt$ and ZH thresholds, obtaining a sample of $10^9$ polarized $Z$ bosons
would provide a very precise determination of $\sin^2\theta_W$, and the $W$
mass
could be measured very precisely at the
$W^+W^-$ threshold~\cite{DESY-ECFA}. However, a centre-of-mass energy of 2
TeV would be
necessary
to ensure full complementarity to the LHC enabling, e.g., the sparticle
spectrum
in Table 1 to be completed.

The first-generation linear collider will enable detailed studies of the Higgs
boson (or the lightest Higgs boson in the MSSM) to be made. Its mass will be
measured to a few parts in $10^4$, and all its major decay modes will be
measured quite accurately~\cite{Battaglia}. This will enable, e.g., a
Standard-Model Higgs boson
to be distinguished from the lightest MSSM Higgs boson, if the heavier MSSM
Higgs
bosons weigh less than several hundred GeV. Even if the centre-of-mass
energy is
restricted to 1 TeV, most of the weakly-interacting sparticles and Higgs bosons
will still be observed directly, and the many spectroscopic measurements will
permit detailed checks of supersymmetric models~\cite{Blaire}.

If an $e^+e^-$ linear collider gets above the threshold for producing pairs of
supersymmetric particles, it will find a scornucopia of new physics. But how likely
is this to happen? The only argument I know that sets a hard upper limit on the
sparticle mass scale is that the LSP constitutes the cold dark matter in the
Universe~\cite{EGO}. This requirement imposes the upper limit
$\Omega_{\rm CDM}h^2 \lappeq$
0.3. However, the supersymmetric relic density $\Omega_\chi$ rises as $m_\chi$ and
the input sparticle mass scales $m_0$, $m_{1/2}$ increase. Therefore, the region of
the $(m_0,m_{1/2})$ plane where $\Omega_\chi h^2 \lappeq$ 0.3 is bounded, although
it has been stretched to larger $m_{1/2}$ by the recent
realization~\cite{EFO} that $\chi
\tilde\ell$ coannihilation processes can be important.

The question then arises, how much of the supersymmetric dark matter region is
covered by a linear $e^+e^-$ collider with a given $E_{\rm cm}$? The answer we
found~\cite{EGO} was about 60~\% if $E_{\rm cm}$ = 0.5 TeV, 90~\% if
$E_{\rm cm}$ = 1 TeV, and
100~\% if $E_{\rm cm}$ = 1.25 TeV, as seen in Fig.~10. The coannihilation
processes were costly: without them, $E_{\rm cm}$ = 0.5 TeV would have been
sufficient to ``guarantee" the discovery of supersymmetry, as seen in
Fig.~12.

\begin{figure}%10
\begin{center}
\mbox{\epsfig{file=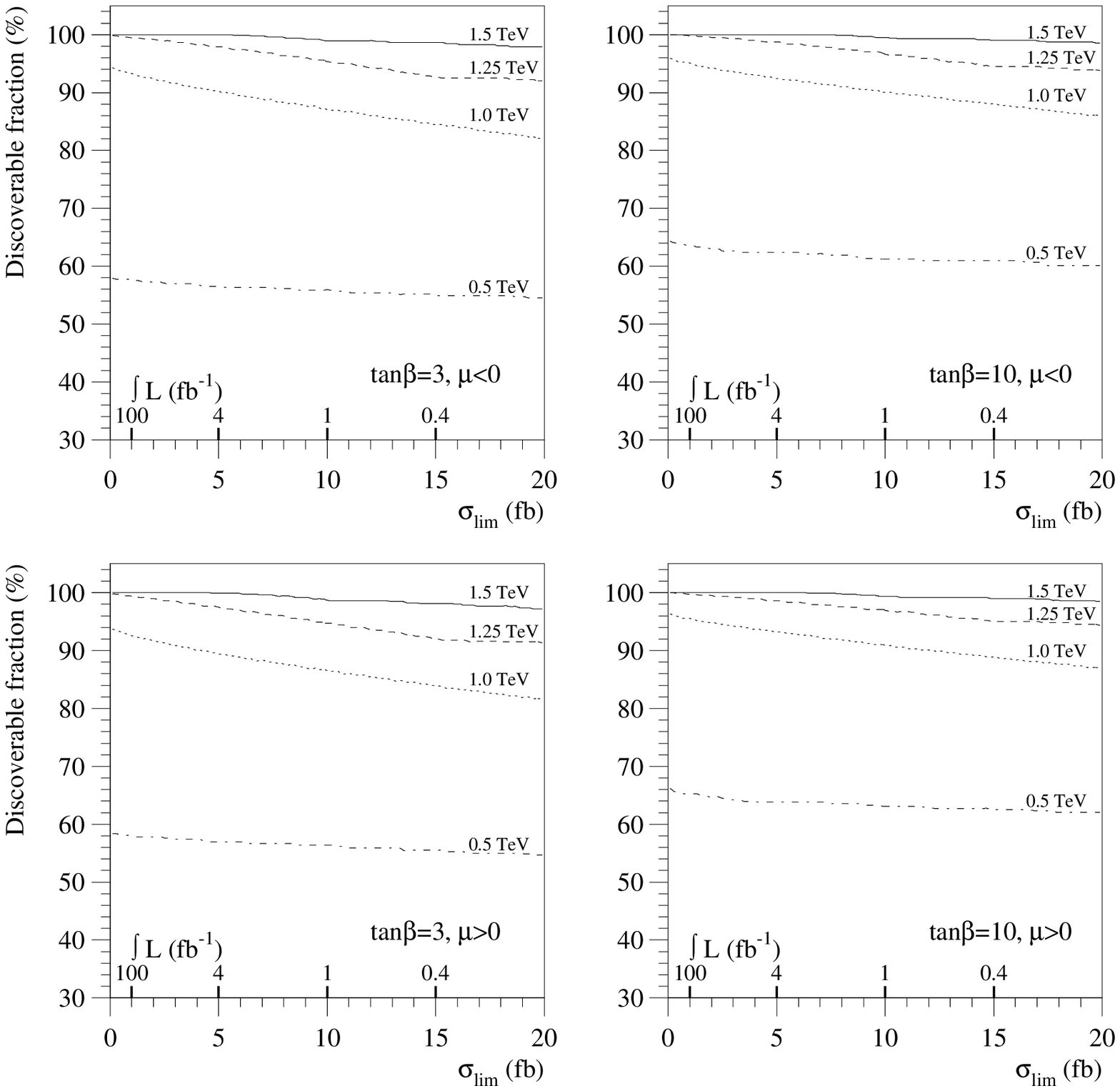,width=8cm}}
\end{center}   
 \captive{\it Percentages of the MSSM parameter space in which the
lightest supersymmetric particle could constitute the cold dark matter
that could be accessed by a linear collider with the indicated
$E_{CM}$ and integrated luminosity~\cite{EGO}.}
\end{figure}

\begin{figure}%11
\begin{center}
\mbox{\epsfig{file=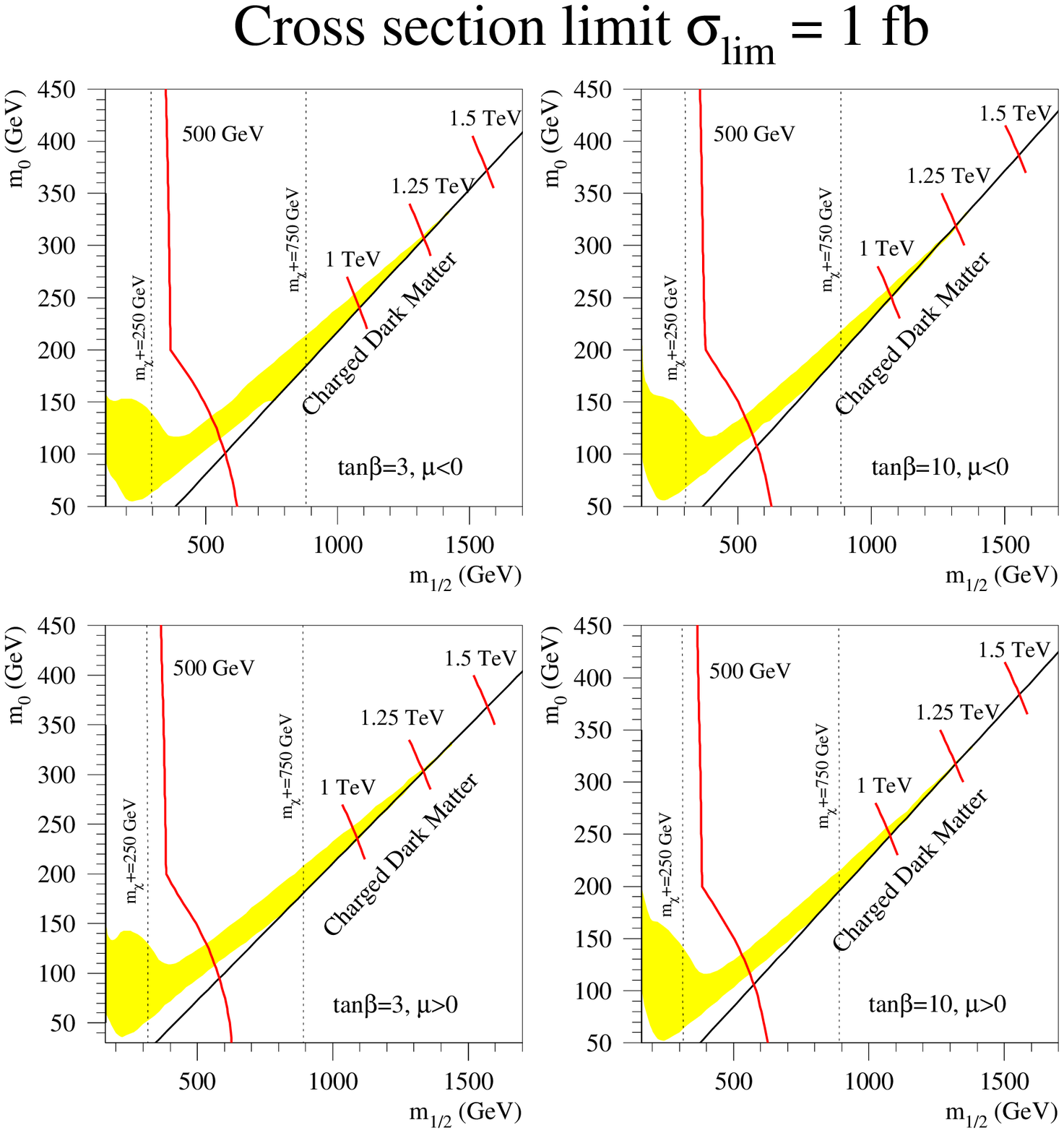,width=8cm}}
\end{center}   
 \captive{\it Illustration of the region of
the MSSM parameter space in which the
lightest supersymmetric particle could constitute the cold dark matter,
compared with the reaches of some 
channels for $e^+ e^-$ annihilation into pairs
of sparticles~\cite{EGO}.}
\end{figure}

It has long been clear to me that physics needs a 1-TeV linear
$e^+e^-$ collider, because of its complementarity to the
LHC~\cite{Jerusalem}. It will be
able to
follow up explorations made with the LHC by making many precision measurements.
As already emphasized, the widest possible energy range is desirable. This
implies that any initial lower-energy phase should be extensible to at
least 1
TeV, and running back in the LEP energy range would also be desirable. For the
rest of this talk, I assume that these physics arguments are sufficiently
strong
that a first-generation 1-TeV linear $e^+e^-$ collider will be built
somewhere.

Nevertheless, there may still be some items on the theoretical wish-list
after the first-generation linear $e^+e^-$ collider. It
would be desirable to have an  accurate direct measurement of the total Higgs
decay width via $s$-channel production, and its mass could be measured much
more
precisely with a muon collider~\cite{Hfact}, as discussed below.
Completing the sparticle
spectrum may require a centre-of-mass energy of 2 TeV or more, as provided by a
second-generation linear $e^+e^-$ collider~\cite{CLIC} or a higher-energy
muon
collider, and
the latter could also produce heavier MSSM Higgs bosons in the direct channel.
Looking further afield, the first glimpse of the 10 TeV energy range could
be
provided by a future larger hadron collider with $E_{cm} \gappeq$ 100
TeV~\cite{VLHC}.

\section{Options for Future Colliders @ CERN}

In mid-1997, the CERN Director-General at the time, Chris Llewellyn Smith
mandated `$\ldots$ a brief written report, $\ldots$, on
possible future
facilities that might be considered at CERN after the LHC'. This should
`$\ldots$ not [be] a major assessment of long-term possibilities'. I would
phrase it as thinking about thinking (about thinking?). The principal options
considered in our report~\cite{EKR} were (i) a next-generation linear
$e^+e^-$
collider with
$E_{cm} \gappeq$ 2 TeV,  based on CLIC technology, (ii) a $\mu^+\mu^-$
collider, ultimately in the multi-TeV $E_{cm}$ range, but perhaps including a
`demonstrator' Higgs factory, and (iii) a future larger hadron collider
(FLHC),
primarily for $pp$ collisions with $E_{cm} \gappeq$ 100 TeV, but perhaps
including options for an $e^+e^-$ top factory and $ep$ collisions in the same
(large) tunnel \footnote{We considered an $ep$ collider in the LEP tunnel to be
already an established CERN option~\cite{LHC-LEP}, and 
in any case not one to be considered a `flagship' project.}.

Starting with the option that we considered least appetizing for CERN, 
if only from the point of view of geography~\cite{EKR}, it seems
apparent that a luminosity of at least 10$^{35}$ cm$^{-2}$s$^{-1}$ would be
required to reap full benefit from a FLHC, perhaps even $10^{36}$
cm$^{-2}$s$^{-1}$ if
$E_{cm} \sim$ 200 TeV. This would pose very severe radiation problems for the
detectors, but such a machine could provide the opportunity to explore the
decade
of mass between 1 and 10 TeV, which history suggests would be a priority after
the LHC.

\section{CLIC}

The default option for the next major project in CERN's future is
probably CLIC, whose physics was first studied
in~\cite{LaThuile}, where its complementarity to the LHC was stressed.
See, in particular, the contributions by Altarelli (p.36), Froidevaux
(p.61), Pauss and myself (p.80), and the review by Amaldi (p.323)
in~\cite{LaThuile}. A study group is now starting to take a further
look at the simulation of benchmark process for CLIC~\cite{BS}.
A preliminary list of key physics processes to be studied is given
in Table 2.

\begin{center}
{Table~2: {\it Examples of Benchmark Physics Processes for
CLIC~\cite{BS}.}}\\

\vspace{0.25cm}

\begin{tabular}{|l||c|c|}
\hline
 & & \\
Physics & Main    & CLIC    \\
Channel & Signature & Challenges\\
 & &\\
\hline \hline
 & & \\
Heavy Higgs $H^0$ & $M_{jj}$ & $L$ \\
Strong Symmetry Breaking & & $\gamma\gamma$ bkg, Hermeticity \\
Anomalous $W$ couplings & & $\gamma\gamma$ bkg\\
$e \bar \nu W$ & & $\gamma\gamma$ bkg \\
 & & \\
\hline
 & & \\
Heavy ($H^0$, $A^0$, $H^{\pm}$) Higgses & $M_{jj}$,
$b$-tag & $\frac{dL}{d{\sqrt{s}}}$, Pairs, $R_{beam~pipe}$\\
SUSY~$\tilde g$, $\tilde q$  & $E_{miss}$, tags &
$\gamma\gamma$ bkg, Hermeticity, $R_{beam~pipe}$\\
~~~~~~~~~$\tilde \chi^+ \tilde \chi^-$  & $E_{miss}$ & $\gamma\gamma$ bkg,
Hermeticity\\
~~~~~~~~~$\tilde \ell^+ \tilde \ell^-$  & $E_{miss}$, $\ell$ &
$\frac{dL}{d{\sqrt{s}}}$, Hermeticity\\
 & & \\
\hline
 & & \\
Contact interactions &  & $\sqrt{s}$, $L$\\
$Z^{'}$ (M $\simeq$ 3~TeV) & Direct $M_{jj, \ell \ell}$
& $\frac{dL}{d{\sqrt{s}}}$, $L$ \\
$Z^{'}$ (M $>$ 5~TeV) & $A_{FB}$, $\sigma_{f \bar f}$ & Pairs, Mask,
$R_{beam~pipe}$  \\
Extra Dimensions  & KK, Indirect & $\gamma\gamma$ bkg, $\mu$ bkg \\
 & & \\
\hline
\end{tabular}
\end{center}

The CLIC two-beam high-energy $e^+e^-$ collider scheme ~\cite{CLIC},
has been used to develop
parameter sets for $E_{cm}$ = 3 and 5 TeV. The
central aim
is a cost-effective, affordable strategy for such a higher-energy linear
collider, since the key CLIC advantages of a high accelerating gradient
and (relatively) simple
components are not needed for a first-generation $E_{cm} \lappeq$ 1 TeV linear
collider. Two CLIC test facilities have already been built and operated
successfully, CTF1 and CTF2~\cite{CLIC}. However, the need for at least
two more demonstrator projects is foreseen
before construction of CLIC itself can be envisaged. These are CTF3 in the
years
2000 to 2005, to demonstrate the acceleration potential in a 0.5 GeV
machine, and
then CLIC1 in the years 2005 to 2009, which should attain 75
GeV~\cite{CLIC}. Recall also that no major
capital investment money will become available at CERN before 2009, because
of the
LHC payment schedule. For both the reasons in the two previous sentences,
CLIC is
necessarily on a longer time scale than that proposed for first-generation
linear
collider projects such as TESLA, the JLC or the NLC.

A CERN geological study has indicated that the tunnel for a linear collider
$\sim$ 30 km long could be excavated parallel to the Jura, entirely in suitable
molasse rock: similar conclusions were reached in a study conducted for
Swissmetro (the group that proposes to build a high-speed underground railway
connecting Geneva and other major Swiss cities)~\cite{EKR}. Also, even a
$E_{cm}$ = 4
TeV $\
\mu^+\mu^-$  collider would fit comfortably within the area bounded by the
existing SPS and LEP/LHC tunnels. On the other hand, it is difficult to see how
even a high-field FLHC with $E_{cm}$ = 100 TeV (which would require a
tunnel
circumference in excess of 100 km) could be accommodated in the
neighbourhood of
CERN.

As far as technological maturity is concerned, even though several hurdles need
to be crossed before the CLIC technology is mature -- for example, the beam
delivery system has hardly been studied -- it may be the closest to mass
shell of
the next-generation collider concepts. The technology required for a FLHC
exists in
principle, but the key problem is to reduce the cost per TeV by an order of
magnitude compared to the LHC. This will require innovative ideas for
tunnelling,
as well as magnets and other machine components~\cite{VLHC}.

\section{Muon Storage Rings}

The most speculative option we considered~\cite{EKR} was a $\mu^+\mu^-$ collider, many of
whose components are at best extrapolations of current technologies, with many
others not existing in any form. Considerable R\&D is required even to
establish
the plausibility of the
$\mu^+\mu^-$ collider concept.
This challenge spurred the formation some years ago in the US of the Muon
Collider Collaboration~\cite{MCC}, which groups a hundred or more
physicists and engineers
and has proposed R\&D projects, notably on ionization
cooling~\cite{MUCOOL}. Until recently,
there was little activity in Europe on muon colliders, although some individual
CERN staff members worked with the Muon Collider Collaboration. This disparity
led RECFA to commission in 1998 a prospective study of $\mu^+\mu^-$ colliders,
whose brief was to specify the physics case, to identify areas requiring
R\&D, and
look for potential European resources outside CERN and DESY.

The corresponding report~\cite{MCYB} produced in early 1999 proposed a
three-step scenario
for physics with  muon storage rings at CERN, illustrated in Fig.~12. The
first
step would be a
$\nu$ factory~\cite{nufact}, in which an intense proton source would be
used to produce
muons,
that would be captured and then cooled by a limited factor, before being
accelerated and stored in a ring and allowed to decay, without being brought
into collision. Such a $\nu$ factory had not been considered
in~\cite{EKR}: the physics interest in such a machine had been amplified
in the mean time, in particular by the emerging evidence for atmospheric
neutrino oscillations. The big advantages over a
conventional $\nu$ beam produced directly by hadronic decays are that the
$\nu$ beams produced by $\mu$ decay would have known fluxes, flavours, charges
and energy spectra, and would  comprise equal numbers of $\nu_\mu$ and
$\bar\nu_e$ (or $\bar\nu_\mu$ and $\nu_e$). Such a $\nu$ factory would
surely
be the `ultimate weapon' for $\nu$ oscillation studies. 
One of the most enticing possibilities is the search for CP 
and/or T violation;
\beq
A_{CP} \equiv
{P(\nu_\mu\rightarrow\nu_e) -
P(\bar\nu_\mu\rightarrow\bar\nu_e)\over
P(\nu_\mu\rightarrow\nu_e) +
P(\bar\nu_\mu\rightarrow\bar\nu_e)}~,
\quad\quad A_T\equiv
{P(\nu_\mu\rightarrow\nu_e) -
P(\nu_e\rightarrow\nu_\mu) \over
P(\nu_\mu\rightarrow\nu_e) +
P(\nu_e\rightarrow\nu_\mu)}
\label{three}
\eeq
which becomes
feasible if the LMA MSW solution to the solar neutrino deficit, as seen in
Fig.~13. This could be followed
by a second step (or steps), namely a Higgs factory (or
factories)~\cite{Hfact}, which
could
measure accurately the mass, width and other properties of a Standard Model
Higgs via its direct $s$-channel production, and thus distinguish between it
and the lightest Higgs in the MSSM, strongly constraining its parameter
space in
the latter case. A second factory operating on the adjacent peaks of the other
neutral $H$ and $A$ Higgs bosons of the MSSM would also be interesting,
as seen in Fig.~14, possibly
opening a novel window on CP violation in the Higgs sector. The third step
would
be a high-energy frontier $\mu^+\mu^-$ collider. Its advantages over an
$e^+e^-$
collider would include superior beam-energy resolution and
calibration~\cite{MCYB}, whereas
an $e^+e^-$ collider such as CLIC would also offer beam polarization and the
possibilities of
$e\gamma, e^-e^-$ and $\gamma\gamma$ collisions.

\begin{figure}%12
\begin{center}
\mbox{\epsfig{file=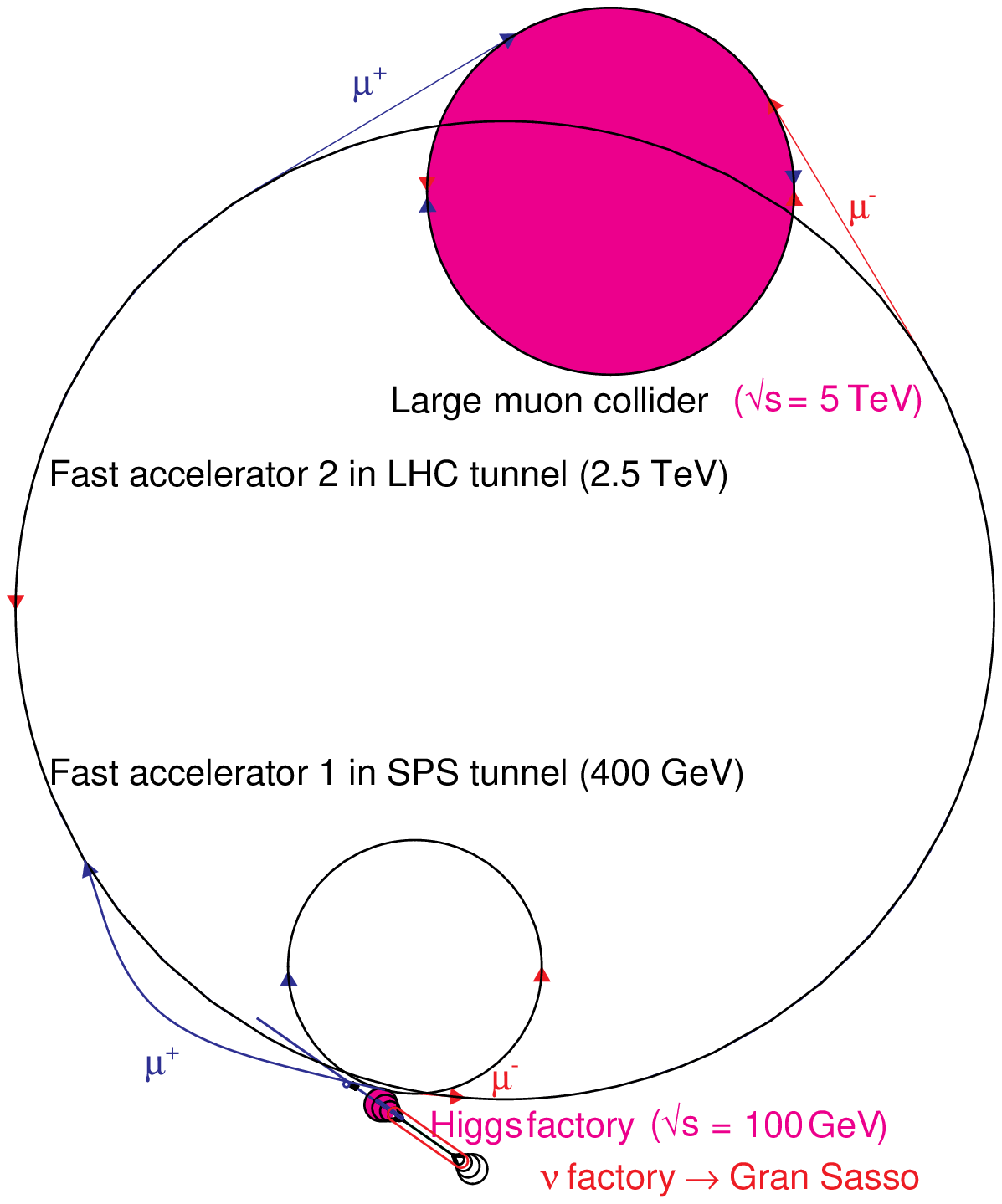,width=8cm}}
\end{center}
 \captive{\it Schematic layout of a possible three-step neutrino
storage ring complex at CERN, including a $\nu$ factory,
a Higgs factory and a possible high-energy frontier muon
collider~\cite{MCYB}.}
\end{figure}

\begin{figure}%13
\begin{center}
\mbox{\epsfig{file=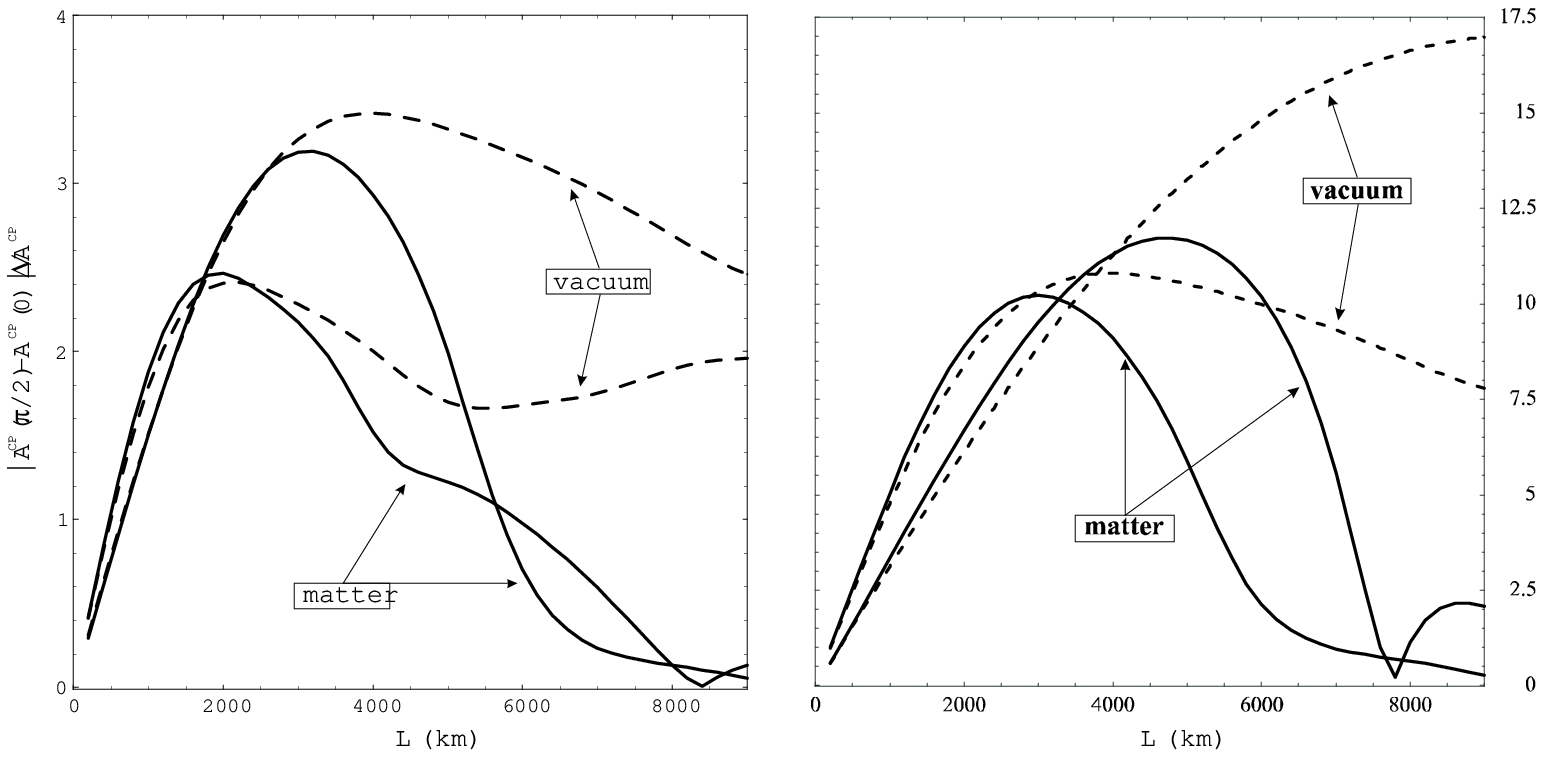,width=14.5cm}}
\end{center}
\captive{\it Significance of the observation of a CP-violating 
asymmetry $A_{CP}$ with $2 \times 10^{20}$ (left panel) or $2 \times
10^{21}$ neutrinos (right panel), including (solid lines) or discarding
(dashed
lines) matter effects, for the mixing parameters described
in~\cite{Donini}.} 
\end{figure}

\begin{figure}%14
\begin{center}
\mbox{\epsfig{file=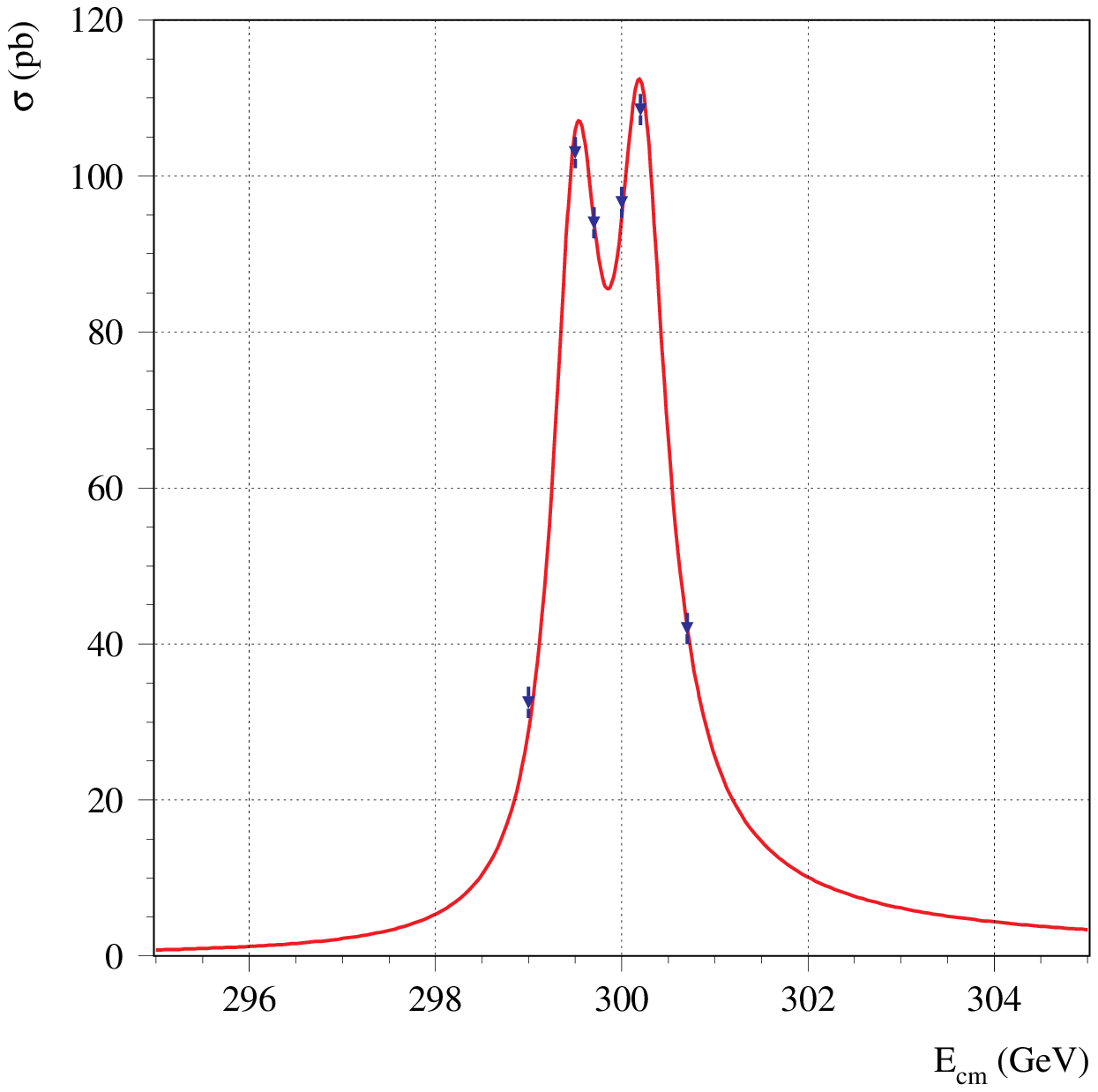,width=8cm}}
\end{center}
\captive{\it Simulated measurements of the direct-channel
production of the heavier MSSM Higgs bosons $(h, A)$ at a Higgs
factory, assuming $m_A = 300$~GeV, tan$\beta = 10$, 
25~pb$^{-1}$ of integrated luminosity per point,
and a beam-energy spread
of $3 \times 10^{-5}$~\cite{MCYB}.}
\end{figure}

Other interesting particle physics~\cite{AIP} would also be possible with
the intense
proton
driver needed for a $\nu$ factory. For example, it might be possible to improve
by several orders of magnitude the current upper limits on
charged-lepton-flavour
violation in the processes $\mu\rightarrow e\gamma$, $\mu\rightarrow 3e$ and
$\mu Z\rightarrow eZ$. Such experiments could explore the range of
interest to
supersymmetric GUT models of $\nu$ oscillations, as seen in
Fig.~15~\cite{EGLLN}. 
\begin{figure}%15
\begin{center}
\mbox{\epsfig{file=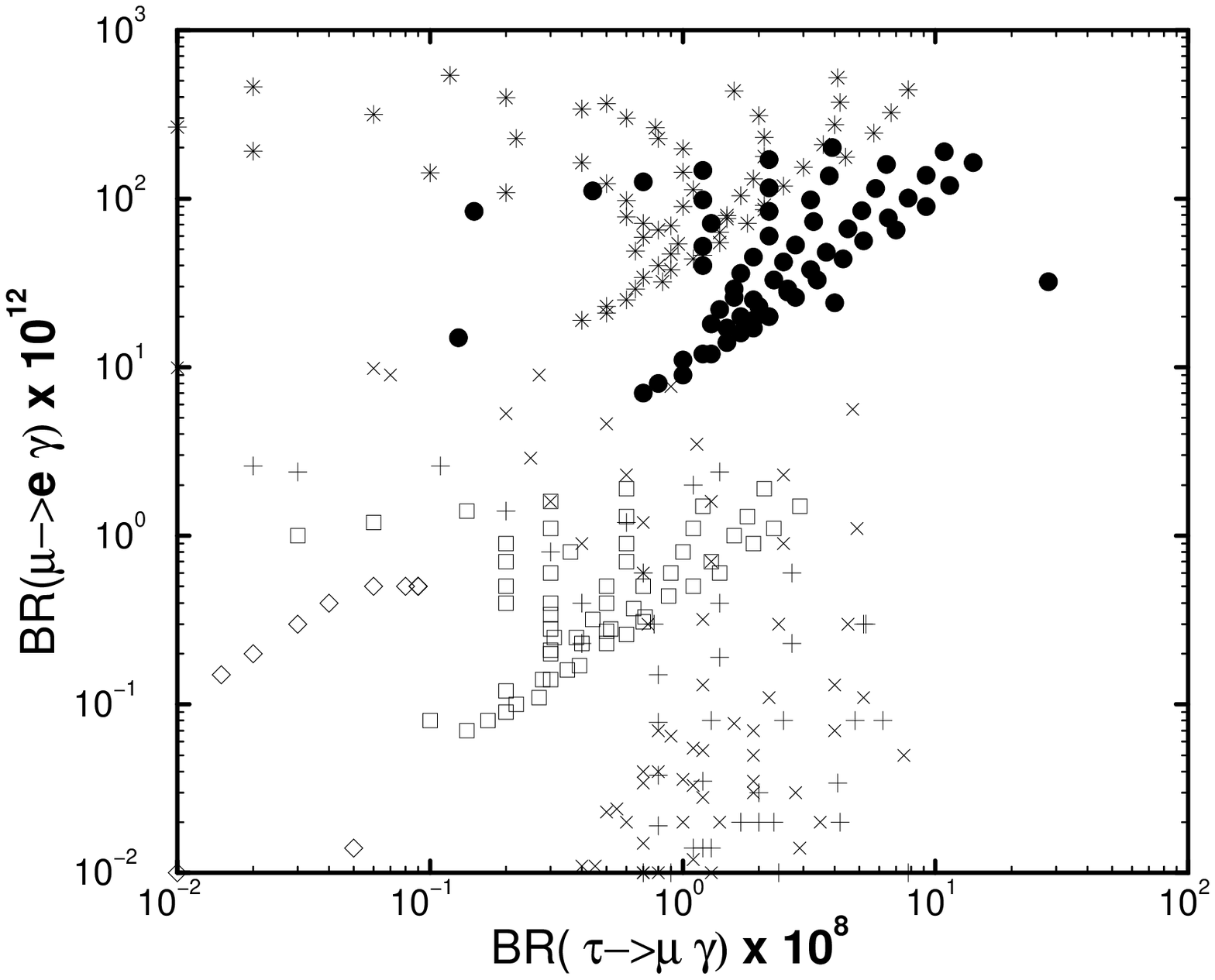,width=8cm}}
\end{center}
\captive{\it Rates for $\mu \rightarrow e \gamma$ and
$\tau \rightarrow \mu \gamma$ decay in some generic
supersymmetric GUT models inspired by the Super-Kamiokande
data on neutrino oscillations, showing opportunities both
for intense $\mu$ beams and for intense $\tau$ sources, such as the
LHC~\cite{EGLLN}.}
\end{figure}

\section{The High-Energy Frontier}

A lepton collider with several TeV of centre-of-mass energy would have a
physics
reach extending beyond the LHC in many respects. What might be interesting
physics at that time, and how would high-energy $e^+e^-$ (e.g., CLIC) and
$\mu^+\mu^-$ colliders compare`\cite{BS}? Their effective mass
reaches may be assumed
to be
similar: $E_{cm}$ for CLIC might be limited for both financial and technical
reasons, and $E_{cm}$ for a $\mu^+\mu^-$ collider might be limited by the
danger
of neutrino radiation~\cite{nurad}, as discussed later.

The difference between the lepton flavours might play a role in some physics
processes, for example in the context of $R$-violating
supersymmetry~\cite{MCYB}, where
$e^+e^-$ and $\mu^+\mu^-$ colliders are sensitive to different couplings.
We have
already seen how the larger Higgs-$\mu^+\mu^-$ coupling could confer advantages
on a lower-energy $\mu^+\mu^-$ collider. The same would be true of a
higher-energy $\mu^+\mu^-$ collider if, for example, $m_A$ were very large,
i.e.,
above 2 TeV. The smaller energy spread and better energy calibration of  a
higher-energy $\mu^+\mu^-$ collider could also be interesting, for example for
threshold measurements. One example studied~\cite{MCYB} was the reaction
$\mu^+\mu^-\rightarrow\chi^+\chi^-$,  where the threshold cross section is much
more sensitive to $m_{\chi^\pm}$ than is $e^+e^- \rightarrow 
\chi^+\chi^-$.
There could also be some advantage in the study of narrow
resonances, as
might occur in some models of strongly-coupled Higgs sectors and/or extra
dimensions~\cite{MCYB}.

On the
other hand, there
are some instances where the
availability of
$e\gamma$,
$\gamma\gamma$ and $e^-e^-$ collisions with an $e^+e^-$  collider could be
advantageous~\cite{BS}. Table~3~\cite{MCYB} lists some relevant physics
topics,
summarizes the
principal capabilities of high-energy $\mu^+\mu^-$ and $e^+e^-$ colliders and
compares them with the LHC. Noted specifically are examples where the energy
precision (E) or flavour non-universality (F) would be advantageous for a
$\mu^+\mu^-$ collider, and where the availability of $e\gamma$ and/or
$\gamma\gamma$ collisions ($\gamma$) or beam polarization (P) would favour an
$e^+e^-$ collider. It should also be commented that the experimental
environment
at a high-energy $\mu^+\mu^-$ collider is likely to be far more difficult
than a
CLIC. There is no way to prevent off-momentum $\mu^\pm$ passing through the
detector, though it should be possible to shield out the $e^\pm$ from $\mu^\pm$
decays.
\begin{table}
\caption{\it A comparison of some of the capabilities of high-energy
colliders, including the LHC, a second-generation linear $e^+ e^-$
collider and a $\mu^+ \mu^-$ collider at the high-energy frontier. for the
latter two cases, we note instances where photon beams ($\gamma$),
polarization (P), flavour non-universality (F) and energy calibration and
resolution (E) might be advantages.} 
\begin{center} \begin{tabular}{|l|l|l|l|} \hline
Physics topics & LHC & $e^+e^-$ & $\mu^+\mu^-$ \\ \hline
Supersymmetry  &&& \\
~~Heavy Higgses H, A & X? & ?:$\gamma$ & Y:F,E \\
~~Sfermions & $\tilde q$ & $\tilde\ell$& $\tilde\ell$: F \\
~~Charginos & X? & Y: P & Y: F,E \\
~~$R$ Violation & $\tilde q$ decays & $\lambda_{1ij}$ & $\lambda_{2ij}$: F,E \\
~~SUSY breaking & some & more & detail: F,E \\ \hline
Strong Higgs sector &&& \\
~Continuum & $\lappeq$ 1.5 TeV & $\lappeq$ 2 TeV & $\lappeq$ 2 TeV \\
~~Resonances & scalar, vector & vector, scalar & vector (E), scalar (F) \\ \hline
Extra dimensions &&& \\
~~Missing energy & large $E_T$ & Y & Y : E? \\
~~Resonances& $q^*, g^*$ & $\gamma^*, Z^*, e^*$ & $\gamma^*, Z^*, \mu^*$: E \\ \hline
\end{tabular}
\end{center}
\end{table}

The biggest obstacle to obtaining high energies in $\mu^+\mu^-$ colliders
may be
$\nu$ radiation~\cite{nurad}, which may even become a health hazard at
$E_{cm} \gappeq$
3 TeV.
Neutrinos will radiate in all directions in the plane of the collider ring,
with
particular concentrations in the directions of any straight sections. In
contrast
to a $\nu$ factory, where these should be as long as possible relative to the
arcs,  in a high-energy $\mu^+\mu^-$ collider one would like them to be as
short
as possible. Other strategies for reducing the $\nu$ radiation hazard include
burying it in a deeper tunnel, learning to be more efficient in using muons to
produce collider luminosity, and subtle choices of `ring' geometry.

\section{Present Accelerator R\&D Activities at CERN}

In its current Medium-Term Plan,
the present CERN management has expanded accelerator R\&D activities
at CERN, including work on both linear colliders and high-intensity
proton sources. A larger fraction of the resources available will be
directed
towards CLIC.  It is hoped~\cite{CLIC} to
continue the
previous
successful studies with two successive stages, CTF3 and CLIC1, before
reaching a
stage (after 2008) when CLIC could be built. In parallel, a working group has
recently been charged to map out a strategy for R\&D towards a $\nu$ factory,
including studies of the proton driver, targetry, $\pi$ capture, $\mu$
cooling and
acceleration~\cite{NFWG}.

As a first step, four specific activities have been proposed~\cite{MUG}:

$\bullet$
an experiment to measure $\pi$ production~\cite{HARP}, which could also
constrain calculations of atmospheric neutrino fluxes,

$\bullet$
tests of RF cavities in a radiation environment
with a strong magnetic field,

$\bullet$
measurements of wide-angle muon  scattering,
with a view to better modelling of cooling channels, and

$\bullet$
target studies.

In parallel to these accelerator R\&D activities, there are physics study
groups
for $\nu$ beams and detectors (concentrating on oscillation
experiments)~\cite{Dydak}, on
$\mu^+\mu^-$ colliders~\cite{Janot}, and
on other possible physics with stopped muons, $\nu$ scattering,
etc.~\cite{EG}.
These activities are in parallel to the accelerator and physics working groups at
FNAL~\cite{FNAL},
the  Expression of Interest for  R\&D towards a
$\nu$
factory submitted to the NSF~\cite{NSF}. The next forum 
for comparing ideas will be the second  
international $\nu$ factory workshop  scheduled for
Monterey in May
2000~\cite{Monterey}.

\section{Prospects}

CERN's experimental programme addresses squarely the
fundamental problems of physics beyond the Standard Model listed
in the Introduction. LEP and the LHC address the problem of {\bf Mass}
in their searches for the Higgs boson and supersymmetry. The
problem of {\bf Unification} is addressed by the CNGS project and
potentially by sparticle mass measurements at the LHC. The
problem of {\bf Flavour} is being addressed by the NA48 experiment,
to be followed by CNGS and B experiments at the LHC.

As reviewed in this talk, in addition to its ongoing programme at LEP and
elsewhere, the CNGS project and its core LHC programme, there are clearly
several interesting options for possible accelerators at CERN beyond the
LHC, which may pursue these problems further. Some of these future
possibilities are being studied quite actively, with CLIC as a default
option~\cite{CLIC}. The relative priorities of the various options before
CERN will depend on project developments elsewhere as well as on physics
developments.

In the coming years, there will clearly need to be mutual understanding
and coordination between accelerator laboratories in different regions of
the world, so as to arrive at a suitable distribution of projects. There
is already worldwide interest in linear $e^+e^-$ colliders, and active
discussion of different projects. In a few years' time, a similar stage
may be reached for $\nu$ factories. Global coordination on R\&D is already
underway, and a similarly cooperative approach to siting optimization
would be desirable.  Hopefully, we will eventually see a `World-Wide
Neutrino Web' consisting of an intense proton source in one region feeding
neutrino beams to detectors in different regions -- a true World
Laboratory for $\nu$ Physics. A Eurocentric vision of this concept is
shown in Fig.~16: see~\cite{NSF} for two competing American visions.

To conclude: in addition to preparing the LHC, 
CERN is preparing actively~\cite{CLIC,BS,NFWG,MUG} to play
whatever role seems most interesting and
appropriate in the generation of accelerators following
the LHC.

\begin{figure}%16
\begin{center}
%\mbox{\epsfig{file=EllisLundfig16.eps,width=8cm}}
\mbox{\epsfig{file=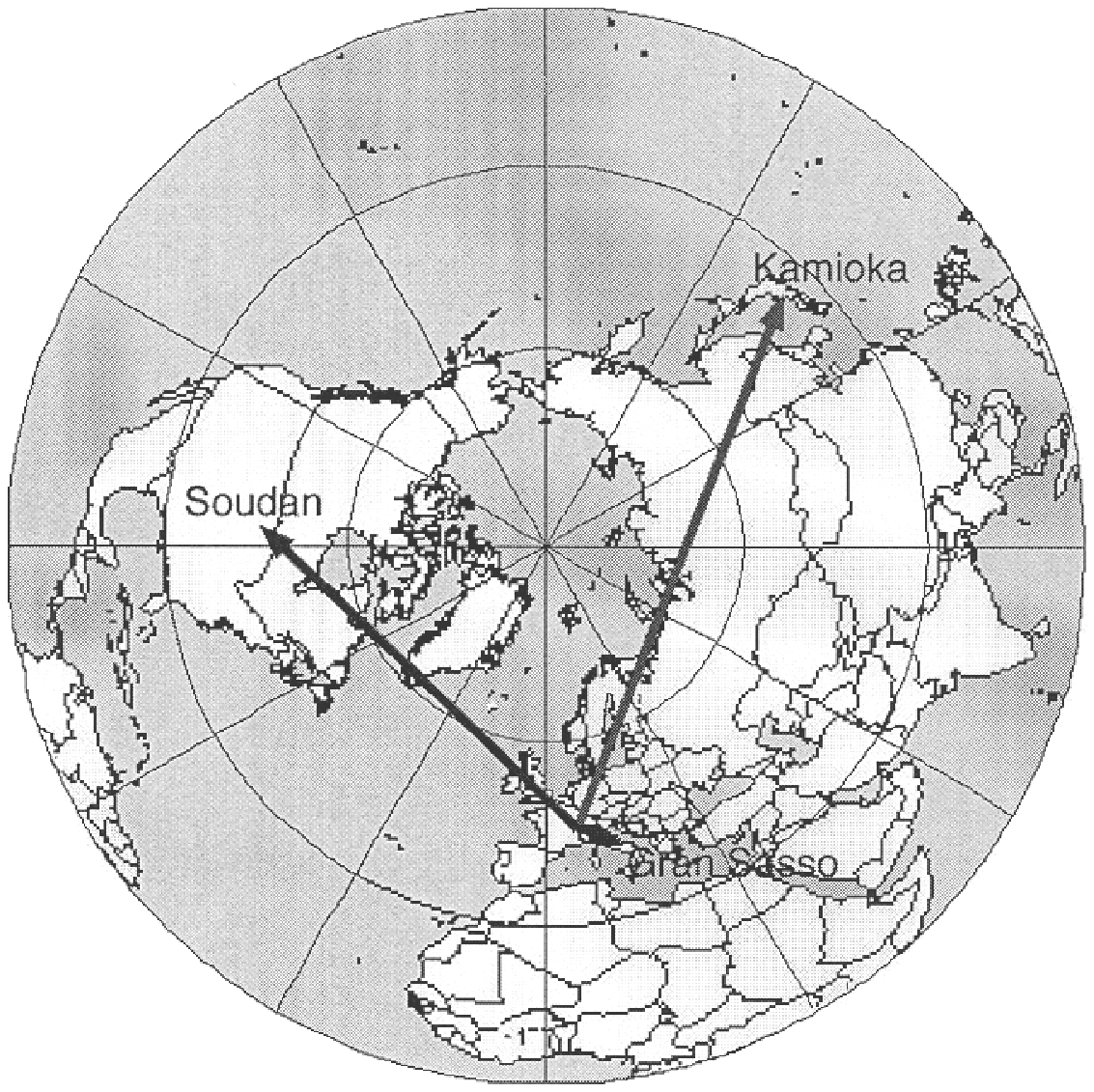,width=8cm}}
\end{center}
\captive{\it A Eurocentric view of the possible World-Wide Neutrino
Web, showing a source at CERN sending $\nu$ beams to the Gran Sasso
laboratory, the Soudan mine and Super-Kamiokande.}
\end{figure}

{\bf Acknowledgements}
{~~}\\
\noindent
It is a pleasure to thank many colleagues for discussions on the
issues raised here, in particular Bruno Autin, Alain Blondel,
Friedrich Dydak, Belen Gavela, Helmut Haseroth, Eberhard Keil,
Luciano Maiani and
Gigi Rolandi. However, I take personal responsibility for the views
expressed here.

\end{document}